\documentstyle[12pt]{article}
\begin{document}
\baselineskip6.8mm
\renewcommand{\theequation}{\thesection.\arabic{equation}}

\vskip 2cm
\title{Stationary Strings and Principal Killing Triads in 2+1  
Gravity}
\author{\\
V. Frolov${}^{*} {}^{1,2,3}$\,
S.Hendy${}^{\dag} {}^{1}$ and A.L.Larsen${}^{||} {}^{1}$}
\maketitle
\noindent
$^{1}${ \em
Theoretical Physics Institute, Department of Physics, \ University of
Alberta, Edmonton, Canada T6G 2J1}
\\ $^{2}${\em CIAR Cosmology Program}
\\ $^{3}${\em P.N.Lebedev Physics Institute,  Leninskii Prospect 53,  
Moscow
117924, Russia}
\begin{abstract}
\baselineskip=1.5em
A new tool for the investigation of $2+1$ dimensional gravity is  
proposed. It is shown that in a stationary  $2+1$ dimensional  
spacetime, the eigenvectors of the covariant derivative of the  
timelike Killing vector form a rigid structure, the {\it principal  
Killing triad}. Two of the triad vectors are null, and in many  
respects they play the role similar to the principal null directions  
in the algebraically special 4-D spacetimes.  It is demonstrated that  
the principal Killing triad can be efficiently used for  
classification and study of stationary $2+1$ spacetimes.

One of the most interesting applications is a study of minimal  
surfaces in a stationary spacetime. A {\it principal Killing surface}  
is defined as a surface formed by Killing trajectories passing  
through a null ray, which is tangent to one of the null vectors of  
the principal Killing triad. We prove that a principal Killing  
surface is minimal if and only if the corresponding null vector is  
geodesic. Furthermore, we prove that if the $2+1$ dimensional
spacetime contains a static limit, then the only regular stationary  
timelike minimal 2-surfaces that cross the static limit, are the  
minimal principal Killing surfaces.

A timelike minimal surface is a solution to the Nambu-Goto equations  
of motion and hence it describes a cosmic string configuration. A  
stationary string interacting with a $2+1$ dimensional rotating black  
hole is discussed in detail.
\end{abstract}

\noindent
PACS numbers: 0420, 0240, 1117\\
\\
Keywords: 2+1 gravity,  Differential Geometry, Cosmic Strings,  Black  
Holes\\
\\
\noindent
$^{*}$Electronic address: frolov@phys.ualberta.ca\\
$^{\dag}$Electronic address: hendy@phys.ualberta.ca\\
$^{||}$Electronic address: alarsen@phys.ualberta.ca
\newpage
\section{Introduction}
\setcounter{equation}{0}
Gravity in $2+1$ dimensions has been intensively investigated in  
recent years,
largely because it provides a relatively simple laboratory for  
testing ideas
developed for the much more complicated problems of realistic $3+1$  
dimensional
gravity; for a recent review see \cite{carlip}.

It is well-known that the Weyl tensor vanishes in $2+1$ dimensions,  
so that the
Riemann tensor can be written explicitly in terms of the Ricci  
tensor. This
means that vacuum solutions to the Einstein equations are locally  
flat; any
deviation from flat $2+1$ dimensional Minkowski spacetime must be  
merely
topological. This is no longer true, of course, when matter sources  
are
admitted. Indeed, there is now a long list of known $2+1$ dimensional
spacetimes with non-trivial geometrical and topological structures;  
for reviews
see  Refs.\cite{barrow,mann}. Furthermore, many of these $2+1$  
dimensional
spacetimes (e.g. some black hole spacetimes) are believed to be able  
to provide
important clues towards a better understanding of "related" $3+1$  
dimensional
spacetimes.

In $2+1$ dimensions there is actually a three-tensor which in some  
respects
plays the same role as the Weyl tensor in higher dimensions; the  
so-called
Cotton tensor \cite{des} (for some early applications, see  for  
instance \cite{york,des1}). A classification of $2+1$
dimensional spacetimes, according to the eigenvalues of the Cotton  
tensor, was
carried out in Ref.\cite{barrow}. In this paper we propose another
classification which works for stationary $2+1$ spacetimes and which  
can be
used instead of the Petrov classification.

One of the aims of the present paper is to demonstrate that in a  
stationary
$2+1$ dimensional spacetime, there is a uniquely defined rigid  
structure, the
{\it principal Killing triad}: provided that the covariant derivative  
of the
timelike Killing vector has a non-zero eigenvalue, the three  
independent
eigenvectors constitute a basis for the spacetime. Up to  
normalization, this
basis is defined uniquely. The triad consists of two linearly  
independent null
vectors $l_\pm,$ and  a unit spacelike vector $m$ orthogonal to them.

We first express the geometrical structures (metric, Killing vector,  
twist,...)
in terms of the principal Killing triad. We then obtain some general  
conditions
for the class of $2+1$ dimensional stationary spacetimes for which  
one or both
of the two principal Killing null eigenvectors are geodesics. It is  
interesting
to notice that most of the $2+1$ dimensional spacetimes of physical  
interest
studied in the literature actually fall in this class (Section 2).

A further goal of this paper is the study of 2 dimensional minimal  
surfaces
embedded in the $2+1$ dimensional spacetimes.  Such surfaces are  
interesting
because they are solutions of the Nambu-Goto equations of motion and  
hence
describe the interaction of a cosmic string with a background  
gravitational
field, in the limit where the cosmic string is infinitely thin.  For  
this
purpose we define a {\it principal Killing surface} as a stationary  
2-surface
formed by Killing trajectories passing through a null ray tangent to  
one of the
null vectors of the principal Killing triad. We prove that a
principal Killing surface is minimal if and only if the corresponding  
null
vector is geodesic (Section 3). Furthermore, we prove a uniqueness  
theorem,
that is, if the $2+1$ dimensional stationary spacetime contains a  
static limit,
then the only regular stationary timelike minimal 2-surfaces that  
cross the
static limit, are the minimal principal Killing surfaces. We also  
discuss
possible applications of this result (Section 4).

Finally, as a special example, we consider the $2+1$ dimensional  
black hole
anti de Sitter spacetime \cite{banados}, using our general formalism.
We find all regular stationary minimal timelike 2-surfaces (string
world-sheets) in this background. Particularly interesting are the  
strings
which cross the static limit. It turns out that the induced geometry  
on the
world-sheets of these configurations
corresponds to global $1+1$ anti de Sitter spacetime. We discuss the  
physical
implications of this result, and we consider the equation determining  
the
propagation of perturbations ({\it stringons}) along these strings  
(Section 5).

\section{The Principal Killing Triad}
\setcounter{equation}{0}

In this section we introduce the notion of the principal Killing  
triad for a
stationary $2+1$ dimensional manifold. We choose spacetime  
coordinates
$x^\mu=(t,x^{i});\;\mu=0,1,2;\;{i}=1,2,$ so that the Killing vector  
$\xi^\mu$
is given by $\xi^\mu=\delta^\mu_t.$ A $2+1$ dimensional stationary  
metric has
the form:
\begin{equation}
g_{\mu\nu}= \pmatrix{
-F & -FA_{i} \cr
-FA_{i} & \frac{H_{ij}}{F}-FA_{i}A_{j} \cr
},
\end{equation}
where $\partial_t F=0,\;\;\partial_t A_{i}=0,\;\;\partial_t H_{ij}=0$  
and
$\xi^\mu\xi_\mu=-F.$
We also assume that  $\xi^\mu$ is timelike $(F>0)$ - at least in some  
open
region of the manifold under consideration.

The Killing equation,
\begin{equation}
\xi_{\mu;\nu}+\xi_{\nu;\mu}=0,
\end{equation}
implies that the tensor $\xi_{\mu;\nu}$ is anti-symmetric. The  
explicit form of
$\xi_{\mu;\nu}$ in the metrics (2.1) can be easily obtained in  
general, see
Appendix A, but it is not important here. Since $\xi_{\mu;\nu}$ is an
anti-symmetric $3\times 3$ matrix, it has eigenvalues  
$(-\kappa,+\kappa,0).$
Denote the  corresponding eigenvectors as
$(l^\mu_+,l^\mu_-,m^\mu)$:
\begin{equation}
\xi_{\mu;\nu}l^\nu_\pm=\mp\kappa l_{\pm\mu},\;\;\;\;\;\;\;\;
\xi_{\mu;\nu}m^\nu=0.
\end{equation}
We would like to stress that a set of three eigenvectors fulfilling  
(2.3) can
be obtained for any (not only timelike) Killing vector of a $2+1$  
dimensional
metric. However, in the rest of this paper we consider the Killing  
vector
$\xi^\mu=\delta^\mu_t$ and the metric (2.1).

In the general case  $\kappa$ is either real or pure imaginary. We  
restrict
ourselves by considering the case when $\kappa=\kappa(x^{i})$ is real  
and
non-zero - at least in some
open region of the manifold under consideration. This turns out to be  
the case
for most $2+1$ dimensional stationary metrics of physical interest,  
and we
shall return to the physical explanation of this, in a moment.
Taking then $\kappa^2>0,$ it follows that
$(l^\mu_+,l^\mu_-,m^\mu)$ are linearly independent vectors, and thus  
provide a
basis for the manifold. We will call this basis the "principal  
Killing triad".
{}From equations (2.3), it  follows that $l_+$ and $l_-$ are null  
vectors,
while $m$ is a spacelike vector orthogonal to both of them:
\begin{equation}
g_{\mu\nu}l^\mu_\pm l^\nu_\pm=0,\;\;\;\;\;\;\;\;\;\;
g_{\mu\nu}l^\mu_\pm  m^\nu=0.
\end{equation}
Without loss of generality, we normalize the principal Killing triad  
so that:
\begin{equation}
g_{\mu\nu}l^\mu_\pm \xi^\nu=-1,\;\;\;\;\;\;\;\;
g_{\mu\nu}m^\mu m^\nu=1.
\end{equation}
With this normalization, the principal Killing triad is uniquely  
defined for a
generic stationary $2+1$ dimensional metric, assuming only that the  
eigenvalue
$\kappa$ is real and non-zero.

The Killing vector $\xi$ takes the following form when expressed in  
the basis
of the principal Killing triad:
\begin{equation}
\xi^\mu=\frac{-1}{l_+\cdot l_-}(l_+^\mu+l_-^\mu)+(m\cdot\xi)m^\mu.
\end{equation}
By covariant differentiation of the identity $\xi^\mu\xi_\mu=-F,$ and
comparison with equations (2.3)-(2.4), we get:
\begin{equation}
\kappa=-\frac{1}{2}\frac{dF}{dx^\nu}l^\nu_+=
\frac{1}{2}\frac{dF}{dx^\nu}l^\nu_-
\end{equation}

The anti-symmetric tensor $\xi_{\mu;\nu}$ written in terms of the  
principal
Killing triad reads:
\begin{equation}
\xi_{\mu;\nu}=-\frac{\kappa}{l_+ \cdot l_-}(l_{+\mu}l_{-\nu}-
l_{+\nu}l_{-\mu}),
\end{equation}
while the metric (2.1) is given by:
\begin{equation}
g_{\mu\nu}=m_\mu m_\nu+\frac{l_{+\mu}l_{-\nu}+
l_{+\nu}l_{-\mu}}{l_+ \cdot l_-}.
\end{equation}
The normalization (2.5) results in the following relationship:
\begin{equation}
-F=(m\cdot\xi)^2+\frac{2}{l_+ \cdot l_-}.
\end{equation}

If a metric is stationary, but not static, the
twist $\Omega,$ defined by,
\begin{equation}
\Omega\equiv-\frac{1}{2\xi^2}e^{\mu\nu\rho}\xi_{\mu;\nu}
\xi_\rho,
\end{equation}
does not vanish
(here  $e^{\mu\nu\rho}$ is the anti-symmetric tensor in 3  
dimensions). The
twist $\Omega$ is the angular velocity of the rotation of the local  
Killing
frame. Using  equations (2.6), (2.8) we can write the twist
$\Omega$ as:
\begin{equation}
\Omega=\frac{\kappa}{F}\; m\cdot\xi.
\end{equation}
Thus the scalar product $m\cdot\xi$ is a measure of the "rotation" of  
the
stationary spacetime: a stationary spacetime with non-zero $\kappa$  
is static
if and only if $m\cdot\xi=0.$ The  twist $\Omega$ can also be  
presented as:
\begin{equation}
\Omega^2=\frac{1}{2}\omega_{\mu\nu}\omega^{\mu\nu},
\end{equation}
where $\omega_{\mu\nu}$ is the velocity of rotation of the Killing  
observer:
\begin{equation}
\omega_{\mu\nu}=\frac{1}{\sqrt{-\xi^2}}\left( \xi_{\mu;\nu}-
(\xi_\mu
a_\nu-\xi_\nu a_\mu) \right),
\end{equation}
and $a_\mu$ is the acceleration of the Killing observer:
\begin{equation}
a_\mu=\frac{1}{2}\nabla_\mu \log (-\xi^2)=\frac{-1}{F}\xi^\nu  
\xi_{\nu;\mu}.
\end{equation}
Notice that:
\begin{equation}
\kappa^2=-\frac{1}{2}\xi_{\mu;\nu}\xi^{\mu;\nu},
\end{equation}
and therefore:
\begin{equation}
\kappa^2=F(a_\mu a^\mu-\frac{1}{2}\omega_{\mu\nu}  
\omega^{\mu\nu})=F(a_\mu
a^\mu-\Omega^2).
\end{equation}
Thus $\kappa^2$ is the acceleration squared minus the angular  
velocity squared.
If therefore we restrict ourselves to consider only $\kappa$ real and  
non-zero
(see the comments after equation (2.3)), we only consider (regions  
of)
spacetimes where the acceleration is larger than the angular velocity  
(in
suitable units).

Let us now examine the null vectors of the principal Killing triad in  
more
detail. For a specific metric, they may or may not be geodesics. In  
fact, we
can classify the stationary $2+1$ dimensional metrics according to  
whether
$l_+$ and/or $l_-$ are geodesics or not. Since the principal Killing  
triad
provides a basis on the manifold, we have the general expansion:
\begin{equation}
l^\nu_\pm l_{\pm\mu;\nu}=A_\pm l_{\pm\mu}+B_\pm l_{\mp\mu}+
C_\pm m_\mu.
\end{equation}
By contracting with $l^\mu_\pm$ we find that $B_\pm=0,$ and by  
contracting with
$\xi^\mu$ that
$A_\pm=(m\cdot\xi)C_\pm,$ that is to say:
\begin{equation}
l^\nu_\pm l_{\pm\mu;\nu}=[(m\cdot\xi) l_{\pm\mu}+ m_\mu]C_\pm,
\end{equation}
where $C_\pm$ are given implicitly by:
\begin{equation}
C_\pm=m^\mu l^\nu_\pm l_{\pm\mu;\nu}.
\end{equation}
Explicit expressions for $C_\pm$ can be obtained from equation (2.3):
\begin{equation}
C_\pm=-m^\mu l^\nu_\pm
(\frac{1}{\kappa}\xi_{\mu;\rho}l^\rho_\pm)_{;\nu}
=-\frac{1}{\kappa} m^\mu
l^\nu_\pm l^\rho_\pm\xi_{\mu;\rho;\nu}
=-\frac{1}{\kappa}
R_{\rho\sigma\mu\nu}m^\mu l^\nu_\pm l^\rho_\pm \xi^\sigma,
\end{equation}
where $R_{\mu\nu\rho\sigma}$ is the Riemann tensor of the manifold.  
This
expression can be rewritten further using the expansion of the  
Riemann tensor
in terms of the Ricci
tensor,  valid generally in $3$ dimensions:
\begin{equation}
R_{\mu\nu\rho\sigma}=g_{\mu\rho}R_{\nu\sigma}+
g_{\nu\sigma}R_{\mu\rho}-g_{\nu\rho}R_{\mu\sigma}-
g_{\mu\sigma}R_{\nu\rho}-\frac{R}{2}(g_{\mu\rho}g_{\nu\sigma}-
g_{\mu\sigma}g_{\nu\rho}).
\end{equation}
A congruence of null rays generated by $l_\pm$ is geodesic if and  
only if
$C_\pm$ vanishes, that is if and only if :
\begin{equation}
(m\cdot \xi)R_{\mu\nu}l^\mu_\pm l^\nu_\pm+R_{\mu\nu}l^\mu_\pm  
m^\nu=0.
\end{equation}
If we consider a "physical" spacetime obtained as a solution of the  
Einstein
equations,
\begin{equation}
R_{\mu\nu}-\frac{R}{2}g_{\mu\nu}=T_{\mu\nu},
\end{equation}
with some physically relevant energy-momentum tensor $T_{\mu\nu},$  
the
condition (2.23) is written:
\begin{equation}
T_{\mu\nu}[(m\cdot \xi) l^\mu_\pm l^\nu_\pm+l^\mu_\pm m^\nu]=0.
\end{equation}
It is also possible to rewrite equation (2.23) as a second order  
non-linear
partial differential equation for the functions $(F,  A_{i},  
H_{ij}),$
appearing in the metric (2.1), but we shall not consider this  
equation and its
general solution here. For our purposes, it will be sufficient to  
consider some
families of special solutions to equation (2.23). It happens that  
these
families of solutions cover most of the known $2+1$ dimensional  
stationary
spacetimes of physical interest.

A family of solutions to Einsteins equations which trivially fulfills  
equation
(2.23) is obtained from an energy-momentum tensor of the form:
\begin{equation}
T_{\mu\nu}=f(x^{i})g_{\mu\nu},
\end{equation}
where $f(x^{i})$ is an arbitrary function. Particular examples in  
this family
of $2+1$ dimensional solutions are the de Sitter and anti de Sitter  
spacetimes \cite{des2}.

More generally, consider the following family of metrics:
\begin{equation}
g_{\mu\nu}= \pmatrix{
-F(x) & 0 & -F(x)A_y(x) \cr
0  & \frac{H_{xx}(x)}{F(x)} & 0\cr
-F(x)A_y(x) & 0 & \frac{H_{yy}(x)}{F(x)}-F(x)A^2_y(x)\cr
};\;\;\;\;\;\;\;{x}^\mu= \pmatrix{
t \cr
x \cr
y \cr
}.
\end{equation}
These are a special case of the metric (2.1) where an additional  
spacelike
Killing vector $\partial_y$ is present.  By explicit construction of  
the
principal Killing triad in this family of metrics,  it can be shown  
that
equation (2.23) is automatically fulfilled for {\it arbitrary}  
functions $F(x),
H_{xx}(x), H_{yy}(x)$
if and only if the function $A_y(x)$ is of the form:
\begin{equation}
 A_y(x)=\frac{c_1}{F(x)}+c_2,
\end{equation}
where $c_1, c_2$
are constants.

We now list some known metrics which belong to this class. Consider  
first the
case when $c_1=c_2=0.$  We already mentioned de Sitter and anti de  
Sitter
spacetimes in connection with equation (2.26). When written in static
coordinates, they are in this class. This is also the case for the  
metric
around a static massive point particle \cite{clement}, as well as for  
the metric around a static closed or open string \cite{des3}.
The static solutions obtained by minimal coupling to a massless  
scalar field or
by coupling to a static magnetic field (or a stiff perfect fluid)
\cite{barrow} are also in this class. So is the charged black string  
solution
of Horne and Horowitz \cite{horne}. Now consider the case when  
$c_2=0$ but
$c_1\neq 0.$ In this class we find the metric of a massless spinning  
point
particle \cite{deser}, as well as the metric of the $2+1$ black hole  
anti de
Sitter (BH-ADS) solution of Banados et al \cite{banados}. In both  
these cases
$c_1$ represents the angular momentum. Notice that the  $2+1$ BH-ADS  
solution
is actually also of the form (2.26)
since it is locally (and asymptotically) isometric to $2+1$
dimensional anti de Sitter spacetime. Thus in all these spacetimes  
both null
vectors  $l_\pm$ of the principal Killing triad are geodesics.

In the cases where one or both of $l_\pm$ are geodesics, the  
principal Killing
basis is an important tool for the study of stationary minimal  
2-surfaces
embedded in the higher dimensional stationary spacetime. This was  
demonstrated
in a recent publication \cite{hendy} concerning the $3+1$ dimensional
Kerr-Newman black hole spacetime, using the corresponding principal  
Killing
tetrad.
In $2+1$ dimensions, the analytical computations are much simpler,  
and it is
possible to consider generic spacetimes. The principal physical  
motivation for
the study of timelike minimal 2-surfaces is that such surfaces are  
solutions of
the Nambu-Goto equations of motion, and hence can represent string
configurations in the spacetime under consideration. We thus consider  
the $2+1$
dimensional spacetimes as convenient toy-models for the investigation  
of the
physics of stationary string world-sheets embedded in higher  
dimensions. This
is the topic of the next section.

\section{String World-Sheets}
\setcounter{equation}{0}

Our aim in this section is to consider stationary regular timelike  
minimal
2-surfaces, corresponding to string world-sheets, embedded in  
stationary $2+1$
dimensional spacetimes where at least one of the two null vectors  
$l_\pm$ of
the principal Killing triad is geodesic. For this purpose we begin by
considering the general properties of stationary
timelike 2-surfaces embedded in a $2+1$ dimensional stationary
spacetime.

Let $\Sigma$ be a 2 dimensional timelike surface embedded in a $2+1$
dimensional stationary
spacetime, and let $\xi$ be the corresponding Killing vector  which  
is timelike
- at least in some open region of the spacetime. $\Sigma$ is said to  
be {\it
stationary} if the Killing vector field $\xi$ is everywhere
tangent to it.  For any such surface $\Sigma$ there
exists two  linearly independent null vector fields, say  $l_1$ and  
$l_2,$
tangent to $\Sigma.$
We assume that the integral curves of each of the null vectors  
$l_{1,2}$ form a
congruence that covers $\Sigma,$
i.e. each point $p \in \Sigma$ lies on exactly one of these integral  
curves.

Thus we can construct a stationary timelike surface $\Sigma$ in the  
following
way: consider a null ray $\gamma$ with tangent vector field $l.$  For  
each
point
$p \in \gamma,$ there is precisely
one Killing trajectory with tangent vector $\xi$ that passes through   
it. The
set of Killing trajectories passing through $\gamma$ forms
a stationary 2-surface $\Sigma$. We define $l$ over $\Sigma$ by Lie
propagation along each Killing trajectory. We call $\gamma$ a basic  
ray of
$\Sigma.$ It is easily verified that $l$ remains null when defined in  
this
manner over $\Sigma$.

We can use the Killing time parameter $u$ along Killing trajectories  
and the
affine parameter $\lambda$
along $\gamma$ as coordinates on $\Sigma.$ In these coordinates
$\zeta^A=(u,\lambda)$ one has $x^{\mu}_{,u}=\xi^{\mu}$ and  
$x^{\mu}_{,\lambda}=
l^{\mu},$ and the induced metric $G_{AB}=g_{\mu \nu} x^{\mu}_{,A}
x^{\nu}_{,B}\;\;(A,B,...=0,1)$ is of the form:
\begin{equation}
dS^2 =G_{AB}d\zeta^A d\zeta^B = - F du^2 + 2(\xi \cdot l) du  
d\lambda.
\end{equation}
Now introduce a spacelike vector $n^{\mu}$ normal to the timelike  
2-surface
$\Sigma$:
\begin{equation}
\label{4.1}g_{\mu \nu} n^{\mu} n^{\nu} = 1, \;\;\;\;\;\;\;\;\;\;
g_{\mu \nu} x^{\mu}_{,A} n^{\nu} = 0.
\end{equation}
For this choice of the triad $(\xi, l, n),$
the following completeness relation is satisfied:
\begin{equation}
\label{4.2}g^{\mu \nu} = G^{AB} x^\mu_{,A} x^\nu_{,B} +   n^{\mu}
n^{\nu}.
\end{equation}
The normal vector $n$ spans the vector space normal to the surface at  
a given
point.

The  second fundamental form $\Omega_{AB}$ for $\Sigma$ is defined  
as:
\begin{equation}
\Omega_{AB}  =  g_{\mu\nu}n^\mu x^\rho_{,A}\nabla_\rho x^\nu_{,B}.
\end{equation}
The condition that a surface $\Sigma$ is minimal can be written in  
terms of the
trace of the second fundamental form:
\begin{equation}
\Omega_{A}\hspace*{1mm}^{A} \equiv G^{AB}\Omega_{AB}=0.
\end{equation}
This is equivalent to the Gauss-Weingarten equation, which for the  
line-element
(3.1) takes the explicit form:
\begin{equation}
2 (\xi \cdot l) l^\rho\xi^\mu_{;\rho}+Fl^\rho l^\mu_{;\rho}+
(\xi \cdot l)
l^\rho\frac{\partial}{\partial x^\rho} \left( {F \over \xi \cdot
l}\right)l^\mu=0.
\end{equation}
It is easily verifed that equation (3.6) is invariant under
reparametrizations of $l^{\mu},$ i.e. if $l^\mu$ satisfies (3.6) then  
so
does $g(x) l^\mu,$ for an arbitrary function $g(x).$ Thus without  
loss of
generality we may normalize $l^\mu$ so
that $l \cdot \xi = -1$. Then equation (3.6) becomes:
\begin{equation}
-2l^\rho\xi^\mu_{;\rho}+Fl^\rho
l^\mu_{;\rho}+l^\rho\frac{\partial F}{\partial x^\rho}l^\mu=0.
\end{equation}

Consider a special type of a stationary 2-surface embedded in a  
stationary
$2+1$ dimensional spacetime with principal Killing triad $(l^\mu_+,
l^\mu_-,m^\mu).$  Namely, consider a surface for which  the null  
vector $l$
coincides with one
of the two null vectors $l_\pm.$ We call
such a surface  $\Sigma_{\pm}$  a {\it  principal Killing surface}  
and
$\gamma_{\pm}$ its {\it basic ray}. We shall use indices $\pm$ to  
distinguish
between quantities connected with $\Sigma_{\pm}$.

In what follows we restrict ourselves by considering timelike
principal Killing surfaces. It is motivated by the fact that this  
class of
surfaces is connected with string world-sheets.
By comparing equations (2.3), (2.7) with equation (3.7), we  
immediately get the
following theorem:\\
\\
A timelike principal Killing surface is a minimal surface if and only  
if the
corresponding null vector of the principal Killing triad is geodesic.  
That is,
if and only if the $2+1$ dimensional stationary metric fulfills  
equation
(2.23), for the corresponding null vector.\\
\\
The theorem is valid in the regions where $\xi^2\neq 0$. In that  
case,
$\Sigma_{\pm}$ is a stationary solution of the Nambu-Goto equations.  
However,
it should be stressed that such minimal principal Killing surfaces  
are only
very special
stationary minimal surfaces.  A general stationary string solution  
(minimal
surface) must  be obtained by explicitly solving equation (3.7),  
which is
generally a highly non-trivial problem. However, it turns out that  
the
minimal principal Killing surfaces play a particularly important role  
if the
stationary $2+1$ dimensional spacetime, besides fulfilling equation  
(2.23),
also contains a curve where the Killing vector $\xi$ becomes null.  
This curve
is defined by:
\begin{equation}
F(x^{i})=0,
\end{equation}
and corresponds to a static limit (or a horizon)  in the $2+1$  
dimensional
spacetime. This will be discussed in the following section.

\section{Uniqueness Theorem}
\setcounter{equation}{0}
Consider a stationary $2+1$ dimensional metric, which fulfills  
equation (2.23)
for at least one of the two null vectors of the principal Killing  
triad. We
also assume that there exists a curve where the Killing vector $\xi$  
becomes
null. This curve is called the static limit curve $S_{\mbox{st}}$ and  
it is
defined by equation (3.8). When constructing a stationary timelike  
2-surface
$\Sigma$ (not necessarily minimal) in such a
spacetime, following the procedure outlined in the previous section,  
we always
choose the null vector $l$ to be that of two possible null vector  
fields
$l_{1,2}$ on $\Sigma$ which does not coincide with the Killing vector  
$\xi$ at
$S_{\mbox{st}}.$ In that case, the metric (3.1) is regular on the  
static limit
curve.

We  prove now that the only stationary timelike minimal 2-surfaces
that cross the static limit curve, and are regular in its vicinity,
are the minimal principal Killing surfaces.

Consider a stationary timelike 2-surface $\Sigma$ with the line  
element (3.1).
The condition that the surface is minimal is given by equation (3.7).
Since $l^\rho l^\mu_{;\rho}$ is regular at $S_{\mbox{st}},$ this  
equation on
the
static limit curve ($F=0$) reduces to:
\begin{equation}
(\xi_{\mu;\rho}-\frac{1}{2}\frac{\partial F}{\partial  
x^\rho}l_\mu)l^\rho=0.
\end{equation}
{}From equations
(2.3), (2.7) it follows that  $l\propto l_+$ (or $l\propto l_-$) on  
the static
limit curve.

Now suppose there exists a timelike minimal surface $\Sigma$  
different from
$\Sigma_+.$ On the static limit curve, $\Sigma$ must have $l\propto  
l_+.$ In
the vicinity of the static limit curve, $l$ can have
only small deviations from $l_+.$
{}From the conditions $l\cdot l=0$ and $l\cdot \xi=-1,$ we then get  
the
following general form of $l$ in the vicinity of the static limit  
curve:
\begin{equation}
l^\mu=[1+B(m\cdot \xi)] l_+^\mu+{B}m^\mu+{\cal
O}(B^2),
\end{equation}
for some function $B,$ where we have only kept terms up to first  
order in $B.$
We insert this expression into equation (3.7),
contract with $m_\mu,$ and keep only terms linear in  $B:$
\begin{equation}
-2{m}_\mu
l^\rho\xi^\mu_{;\rho}=0,\;\;\;\;\;\;\;\;\;\;\;\;(\mbox{to all orders  
in}\;B),
\end{equation}
\begin{equation}
{m}_\mu
l^\rho\frac{\partial F}{\partial x^\rho}l^\mu=l^\rho_+
\frac{\partial
F}{\partial x^\rho}B+
{\cal
O}(B^2),
\end{equation}
\begin{equation}
{m}_\mu F\l^\rho
l^\mu_{;\rho}=Fl_+^\rho\frac{\partial B}{\partial x^\rho}+
F{m}^\rho {m}^\mu l_{+\mu;\rho}
B+{\cal O}(B^2).
\end{equation}
Thus altogether:
\begin{equation}
l^\rho_+\frac{\partial{B}}{\partial  
x^\rho}+[\frac{1}{F}l^\rho_+\frac{\partial
F}{\partial x^\rho}+
{m}^\rho {m}^\mu l_{+\mu;\rho}]B+{\cal O}(B^2)=0.
\end{equation}
Notice that the second term in the bracket corresponds to the  
"expansion"
$\theta$ of the null rays \cite{mtw}:
\begin{equation}
{m}^\rho {m}^\mu l_{+\mu;\rho}=l^\mu_{+;\mu}\equiv\theta.
\end{equation}
Since we consider the 2-surface $\Sigma$ to be timelike and regular,  
even near
the static limit curve, the expansion $\theta$ is regular:  $\theta$  
being
divergent at some point $p,$ would imply the existence of a caustic.  
However,
this contradicts the fact that only one integral curve of the  
null-vector $l_+$
passes through $p,$ as follows from the regularity condition, see  
\cite{beem}.
Near the static limit curve the first term in the bracket of equation  
(4.6)
dominates, and it follows that the solution near the static limit  
curve is
approximated by:
\begin{equation}
{B}=\frac{c}{F} ;\;\;\;\;\;\;\;\;\;\;c=\mbox{const.}
\end{equation}
A solution regular near the static limit curve $(F\rightarrow 0)$
can therefore only be obtained for $c=0,$ which implies that $B=0.$
Thus we have shown that $\Sigma$ is regular and minimal if and only  
if
 $l \propto l_\pm$. This proves the uniqueness theorem:
The only stationary timelike minimal 2-surfaces
that cross the static limit curve $S_{\mbox{st}},$ and are regular in  
its
vicinity,
are the mimimal principal Killing surfaces.

We now discuss possible physical applications of this result.\\
Consider one of the null vectors of the principal Killing triad, say  
$l_+.$
Since $l_+^\mu$ cannot be proportional to $\xi^\mu,$ either $l_+^{x}$  
and/or
$l_+^{y}$ is non-zero (we here use the notation $\mu=(t, x, y)$). Let  
us assume
that (say) $l_+^{x}\neq 0.$ We also restrict ourselves to consider  
only metrics
where an additional Killing vector (say) $\partial_y$ is present  
(that is,
metrics of the form (2.27)). We can then introduce generalized
Eddington-Finkelstein coordinates in the following way:
\begin{equation}
dx^\mu_+=(A_+)^\mu\;_\nu dx^\nu,
\end{equation}
where the matrix $(A_+)^\mu\;_\nu$ is defined by:
\begin{equation}
(A_+)^\mu\;_\nu\equiv \pmatrix{
1 & -l_+^{t}/l_+^{x} & 0 \cr
0 & -1/l_+^{x} & 0 \cr
0 &  l_+^{y}/l_+^{x} & - 1 \cr
},
\end{equation}
that is:
\begin{equation}
dt_+=dt-\frac{l_+^{t}}{l_+^{x}} dx,
\end{equation}
\begin{equation}
dx_+=-\frac{dx}{l_+^{x}},
\end{equation}
\begin{equation}
dy_+=-dy+ \frac{l_+^{y}}{l_+^x} dx.
\end{equation}
In a similar way we can  introduce generalized Eddington-Finkelstein
coordinates corresponding to $l_-.$

We have shown that any stationary timelike minimal 2-surface that  
crosses the
static limit curve in the stationary $2+1$ dimensional spacetime  
(assumed to
fulfill equation (2.23)) and remains regular in its vicinity, is a  
minimal
principal Killing surface. For the principal Killing surface  
$\Sigma_+$ we have
$\partial x^\mu/\partial\lambda=l_+^\mu$ (up to a constant factor).   
We can
choose the affine parameter $\lambda$ such that:
\begin{equation}
dx^\mu=- l_+^\mu d\lambda+\xi^\mu du,
\end{equation}
that is:
\begin{equation}
dt=- l_+^{t} d\lambda +du,
\end{equation}
\begin{equation}
dx=- l_+^{x} d\lambda,
\end{equation}
\begin{equation}
dy=- l_+^{y} d\lambda.
\end{equation}
By comparing Equations (4.11)-(4.13) with equations (4.15)-(4.17), we  
find:
\begin{equation}
dx_+=d\lambda,
\end{equation}
\begin{equation}
dy_+=0,
\end{equation}
\begin{equation}
dt_+=du.
\end{equation}
Thus a minimal principal Killing surface  corresponds to $y_+=$  
const.
Physically it means that a stationary string  can only cross the  
static limit
curve in very special ways, namely along the direction
$y_+=$ const.
The above consideration certainly is also valid for $\Sigma_-.$

We can then write down the induced line-element on the world-sheets  
of
$\Sigma_\pm$ in terms of the generalized Eddington-Finkelstein  
coordinates
using equations (2.9), (2.10), (4.14) and (4.18):
\begin{equation}
dS^2=-F dt_\pm^2\pm 2dt_\pm dx_\pm.
\end{equation}

It is an interesting observation that the static limit curve of the
$2+1$ dimensional spacetime corresponds to a horizon on the string  
world-sheet,
as follows from equations (3.8), (4.21), {\it even} if it does not  
correspond
to a horizon in the
$2+1$ dimensional spacetime \cite{FrAl:95}. This may lead to very  
interesting
causal properties \cite{hendy}, as will be discussed in more detail  
in the next
section.
\vskip 12pt
\hspace*{-6mm}In the general case a minimal surface obtained by a  
small
perturbation of a minimal principal Killing surface is not  
stationary.  A
general transverse perturbation about a background Nambu-Goto string
world-sheet embedded in $2+1$ dimensions can be written as :
\begin{equation}
\delta x^{\mu} = \Phi n^{\mu},
\end{equation}
where the normal vector $n^\mu$ is defined by equations (3.2).
The equation defining the perturbation $\Phi$ follows from the
following effective action \cite{FrAl:94} (see also  
\cite{guven,carter}):
\begin{equation}
{\cal S}_{\mbox{eff.}}=\int_{}^{}
d^2\zeta\sqrt{-G}\;\Phi\left\{G^{AB}
\nabla_A
\nabla_B+
{\cal V}\right\}\Phi,
\end{equation}
where ${\cal V}$ is a scalar potential defined as:
\begin{equation}
{\cal V}\equiv\Omega_{AB}\Omega^{AB}-
G^{AB}x^\mu_{,A}x^\nu_{,B}R_{\mu\rho\sigma\nu}n^\rho
n^\sigma.
\end{equation}
The equation describing the propagation of perturbations on the  
world-sheet
background
is then found to be:
\begin{equation}
\left\{\Box+ {\cal V} \right\} \Phi = 0.
\end{equation}
Consider the scalar potential ${\cal V}\equiv\Omega_{AB}\Omega^{AB}-
G^{AB}x^\mu_{,A}x^\nu_{,B}R_{\mu\rho\sigma\nu}n^\rho
n^\sigma.$ It is easily verified that the first term vanishes for the  
minimal
principal Killing surfaces $\Sigma_\pm.$ The second term is rewritten  
using
equations (2.22), (3.3):
\begin{eqnarray}
{\cal V}\hspace*{-2mm}&=&\hspace*{-2mm}R_{\mu\nu}n^\mu
n^\nu=R-G^{AB}R_{\mu\nu}x^\mu_{,A}x^\nu_{,B}\nonumber\\
\hspace*{-2mm}&=&\hspace*{-2mm}R+F R_{\mu\nu}l_\pm^\mu l_\pm^\nu\pm
2R_{\mu\nu}\xi^\mu l_\pm^\nu.
\end{eqnarray}
The equation (4.25) of propagation of string perturbations  
(stringons) can be
used to  study the stability of the string configurations. One can  
also expect
that if a stationary cosmic string crosses the static limit curve,  
the Hawking
mechanism will provide thermal excitation of quantum stringons.

In Section 5 we shall consider in more detail the equation (4.25) in  
the case
of the $2+1$ dimensional black hole anti de Sitter spacetime.

\section{The 2+1 BH-ADS Spacetime}
\setcounter{equation}{0}
As an important example, we now consider the $2+1$ dimensional BH-ADS  
spacetime
in more detail, using the formalism and tools developed in the  
previous
sections.
This spacetime is often used as a 2+1 dimensional toy model of a  
rotating black
hole.
The metric of the $2+1$ dimensional BH-ADS spacetime is given by
\cite{banados}:
\begin{equation}
ds^2=(\frac{J^2}{4r^2}-\Delta)dt^2+\frac{dr^2}{\Delta}-Jdt dr+r^2  
d\phi^2,
\end{equation}
where:
\begin{equation}
\Delta=\frac{r^2}{l^2}-M+\frac{J^2}{4r^2}.
\end{equation}
Here $M$ represents the mass, $J$ is the angular momentum and the  
cosmological
constant is $\Lambda=-1/l^2.$  The Killing vector  
$\xi^\mu=\delta^\mu_t,$ which
is timelike at infinity, defines the function $F:$\begin{equation}
F=-\xi^2=\frac{r^2}{l^2}-M.
\end{equation}
For $M^2l^2\geq J^2,$ there are two horizons ($g_{rr}=\infty$):
\begin{equation}
r_\pm^2=\frac{Ml^2}{2}\left( 1\pm\sqrt{1-\frac{J^2}{M^2l^2}}\;\right)
\end{equation}
and a static limit curve ($g_{tt}=0$):
\begin{equation}
r^2_{\mbox{st}}=M l^2.
\end{equation}
The static limit curve lies outside the external horizon,  
$r_{\mbox{st}}\geq
r_+,$ so the causal structure is similar to that of the 4-dimensional
Kerr-Newman geometry. Notice that in $2+1$ dimensions there is no  
strong
curvature singularity at $r=0, $  in fact:
\begin{equation}
R_{\mu\nu}=2\Lambda g_{\mu\nu}.
\end{equation}
For more details on the local and global geometry of the BH-ADS  
solution, see
for instance \cite{banados,banados2,welch}.

The principal Killing triad, as defined in Section 2, is given by:
\begin{equation}
l^\mu_\pm=(\frac{1}{\Delta},\;\mp1,\;\frac{J}{2r^2\Delta}),
\;\;\;\;\;\;\;\;\;\;m^\mu=(0,\;0,\;\frac{1}{r}).
\end{equation}
The results of Section 2 imply that $l_\pm$ are null geodesics:
\begin{equation}
l^\mu_\pm l_{\pm\mu}=0,\;\;\;\;\;\;\;\;\;\;l^\mu_\pm  
l^\nu_{\pm;\mu}=0.
\end{equation}

We are interested in the regular stationary timelike minimal  
2-surfaces,
corresponding to string world-sheets, embedded in the spacetime  
(5.1). In the
present case we can solve the string equation (3.7) exactly. First we  
write the
null tangent vector $l$  in the general form:
\begin{equation}
l^\mu=Al^\mu_+ +Bl^\mu_- + Cm^\mu,
\end{equation}
for some functions $(A, B, C).$
The two conditions $l\cdot l=0$ and $l\cdot\xi=-1$ imply:
\begin{equation}
C^2=\frac{4AB}{\Delta},\;\;\;\;\;\;\;\;\;\;A+B+\frac{J}{2r}C=1,
\end{equation}
 and equation (3.7) reduces to:
\begin{equation}
\frac{dC}{dr}+\left( \frac{1}{r}+\frac{1}{F}\frac{dF}{dr}\right)=0.
\end{equation}
It follows that:
\begin{eqnarray}
A\hspace*{-2mm}&=&\hspace*{-2mm}\frac{1}{2}
\left(1-\frac{Jk}{2r^2F}\right)
\pm\frac{1}{2}\sqrt{1-\frac{Jk+k^2}{r^2F}},\nonumber\\
B\hspace*{-2mm}&=&\hspace*{-2mm}\frac{1}{2}
\left(1-\frac{Jk}{2r^2F}\right)\mp
\frac{1}{2}\sqrt{1-\frac{Jk+k^2}{r^2F}},\nonumber\\
C\hspace*{-2mm}&=&\hspace*{-2mm}\frac{k}{rF},
\end{eqnarray}
where $k$ is an integration constant. The null tangent vector $l$ is  
then given
by:
\begin{equation}
l^\mu=\frac{\partial x^\mu}{\partial\lambda}=\left(
\frac{1}{\Delta}(1-\frac{Jk}{2r^2F}),\;
\mp\sqrt{1-\frac{Jk+k^2}{r^2F}},\;
\frac{1}{r^2\Delta}(k+\frac{J}{2})\right).
\end{equation}
Regular solutions exist in two different situations:\\
\\
{\bf I}. If $\;k^2+kJ>0,$ then there is a turning point  
$(dr/d\phi=0)$ outside
the static limit:
\begin{equation}
r_0^2=\frac{1}{2}Ml^2 \left(1+\sqrt{1+\frac{4(k^2+kJ)}
{M^2 l^2}}\;\right).
\end{equation}
That is to say, the strings are of the "hanging string type"  
\cite{zel}, i.e.
with both ends at spatial infinity. In the case $J=0,$ these strings  
were
discussed in detail in \cite{norma}.\\
\\
{\bf II}. If  $k=0,$ these strings correspond to $l=l_\pm,$ and  
according to
the uniqueness theorem of Section 4, they are the
only strings that cross the static limit curve and being regular in  
its
vicinity. These strings are of the "spiralling string type"  
\cite{zel}. \\
\\
If instead $k^2+kJ\leq 0$ and $k\neq 0,$ the tangent vector $l$  
diverges at the
static limit curve $F=0,$ that is, the solution is not regular.

We now consider the spiralling strings $(k=0)$ in more detail.
In the generalized Eddington-Finkelstein coordinates \cite{chan}:
\begin{equation}
dt_\pm=dt\pm\frac{dr}{\Delta},
\end{equation}
\begin{equation}
d\phi_\pm=d\phi\pm\frac{J}{2r^2\Delta}dr,
\end{equation}
c.f. equations (4.9)-(4.13), they correspond to:
\begin{equation}
\phi_\pm={\mbox {const.}}
\end{equation}
The metric (5.1) in the generalized Eddington-Finkelstein coordinates  
is:
\begin{equation}
ds^2=-\Delta dt_\pm^2+\frac{1}{4r^2}[Jdt_\pm-2r^2 d\phi_\pm]^2\pm  
2dt_\pm dr,
\end{equation}
so that the induced metric on the world-sheets $\Sigma_\pm$
becomes:
\begin{equation}
dS^2=-F dt_\pm^2 \pm 2dt_\pm dr.
\end{equation}
It must be stressed that the metrics (5.19) for $\Sigma_+$
and $\Sigma_- $ simply describe different portions of global $1+1$  
dimensional
anti de Sitter spacetime.
Contrary to their $2+1$ dimensional origin, the $2+1$ BH-ADS,
the  principal Killing surfaces $\Sigma_\pm$ are thus purely  
"cosmological" in
nature; the 2-D horizon at $r=\sqrt{M}l$ is merely of Rindler-type.

For $J=0,$ $\Sigma_\pm$ are geodesic surfaces in the
3-D spacetime and they describe the two branches of a geodesically  
complete 2-D
manifold ($1+1$ anti de Sitter spacetime). However, for the generic  
BH-ADS
geometry $(J\neq 0)$, only one of two null basic lines of the  
principal Killing
surface, namely the ray $\gamma_{\pm}$ with tangent vector $l_{\pm}$,   
is
geodesic in the three-dimensional embedding space. The other basic  
null ray
$\gamma'$ is
geodesic in $\Sigma_\pm$ but not in the embedding space. This implies  
that in
general (when $J\neq 0$) the principal Killing surface is not  
geodesic.
Furthermore, it can be shown that $\Sigma_{\pm}$ considered as a 2-D  
manifold
is geodesically incomplete with respect to its null geodesic  
$\gamma'$.

As a consequence of $\Sigma_\pm$ not being geodesic (when $J\neq 0$),   
it is
possible to send causal signals from the interior
of the 2-D horizon to its exterior (as seen by an observer using the
coordinates (5.19)) by exploiting the
extra dimension of the 3-D spacetime.
It is evident that there exist causal lines leaving the ergosphere  
and entering
the exterior of the BH-ADS spacetime. It means that the "interior"  
and
"exterior" of $\Sigma_+$  (as seen by an observer using the  
coordinates (5.19))
can be connected by 3-D causal lines. It can also be shown that  the  
causal
line can be chosen
to be a null geodesic. The argument is similar to the situation in  
the 4-D
Kerr-Newman spacetime \cite{hendy} so we shall not repeat it here.

It is an interesting observation that the horizon of $\Sigma_+$ (as  
seen by an
observer using the coordinates (5.19)) coincides with the static
limit of the 3-D rotating black hole. The 2-D surface gravity, which  
is
proportional to the 2-D temperature, is given by:
\begin{equation}
\kappa^{(2)}=\left.
\frac{1}{2}\frac{dF}{dr}\right|_{r=r_{st}}=
\frac{\sqrt{M}}{l}.
\end{equation}
The surface gravity of the 3-D BH-ADS spacetime is \cite{banados}:
\begin{equation}
\kappa^{(3)}=\frac{r_+^2-r_-^2}
{l^2 r_+},
\end{equation}
where $r_\pm$ are defined in equation (5.4).
It can then be easily shown  that:
\begin{equation}
\kappa^{(2)}\geq\kappa^{(3)}.
\end{equation}
That is to say, the 2-D temperature is higher than the 3-D  
temperature (except
when $J=0$) and it is always positive. Even if the 3-D
black hole is extreme, the 2-D temperature is non-zero. It should be  
stressed
that the temperatures discussed here are the temperatures as measured  
by an
observer situated where the timelike Killing vector is normalized  
such that
$\xi^2=-1,$ \cite{brown} that is at $r=l\sqrt{M+1}.$
\vskip 12pt
\hspace*{-6mm}We close this section with the following remark:

The physical properties of the Principal Killing Surface $\Sigma_+$  
can be
investigated by considering the propagation of perturbations  
(stringons) along
the cosmic string. The general equations determining the  
perturbations were
obtained at the end of Section 4, equations (4.25)-(4.26). For  
$\Sigma_+,$ with
metric given by (5.19), the equation determining the perturbations  
is:
\begin{equation}
\left( \Box-\frac{2}{l^2}
\right)\Phi=0.
\end{equation}
It is convenient to make the following coordinate transformations. We  
first
introduce the coordinates $(\tilde{u}, \tilde{r}):$
\begin{equation}
u=\tilde{u}+\tilde{r},\;\;\;\;\;\;\;\;\;\;r=-\sqrt{M}l\;\mbox{cotanh}
\left(\frac{\sqrt{M}\;\tilde{r}}{l}\right),
\end{equation}
and then $(\lambda,\rho):$
\begin{equation}
\tilde{r}=\sqrt{M}l\;\mbox{sec}\rho\cos\lambda,\;\;\;\;\;\;\;\;\;\;
\tanh\left(  
\frac{\sqrt{M}\tilde{u}}{l}\right)=\frac{\sin\rho}{\sin\lambda},
\end{equation}
so that the line element (5.19) takes the standard $1+1$ anti de  
Sitter form:
\begin{equation}
dS^2=l^2\mbox{sec}^2\rho\left( -d\lambda^2+d\rho^2\right),
\end{equation}
and the d'Alambertian is given by:
\begin{equation}
\Box=\frac{\cos^2\rho}{l^2}\left(-\partial_{\lambda}^2+
\partial_{\rho}^2\right).
\end{equation}
In this formalism, the classical and quantum aspects of equation
(5.23) have been discussed in the literature, so we shall not go into  
the
details here. We refer the interested reader to Ref. \cite{sakai},  
and
references given therein.
\section*{Acknowledgements}
\setcounter{equation}{0}
The
work of V.F. and A.L.L  was supported by NSERC, while the work by  
S.H. was
supported by the Canadian Commonwealth Scholarship and Fellowship  
Program.

\appendix
\section{Explicit form of $\xi_{\mu;\nu}$}
\setcounter{equation}{0}
Using the metric (2.1), it is straightforward to compute the  
Christoffel
symbols in terms of $(F, A_{i}, H_{ij}).$ It can then be shown that  
the
anti-symmetric tensor $\xi_{\mu;\nu}$ takes the form:
\begin{equation}
\xi_{\mu;\nu}= \pmatrix{
0 & -\frac{1}{2}F_{,i} \cr
\frac{1}{2}F_{,i} & (FA_{[j})_{,i]} \cr
}.
\end{equation}
The eigenvalues $\lambda$ are obtained from the equation:
\begin{equation}
\mbox{det}(\xi_{\mu;\nu}-\lambda g_{\mu\nu})=0,
\end{equation}
and leads to:
\begin{equation}
\lambda=0\;\;\;\;\;\mbox{or}\;\;\;\;\;\lambda=\mp
\frac{1}{2}[H^{ij}F_{,i}F_{,j}-F^4(e^{ij}A_{i,j})^2]^{1/2}
\equiv\mp\kappa.
\end{equation}
Here $H^{ij}$ and $e^{ij}$ are defined by:
\begin{equation}
H_{ik}H^{jk}=\delta_i^j,\;\;\;\;\;\;\;\;\;\;e^{ij}=\frac{\epsilon^{ij} 
}
{\sqrt{\mbox{det}(H)}},
\end{equation}
where $\epsilon^{ij}$ is the Levi-Civita symbol in two dimensions.
Explicit expressions for the linearly independent eigenvectors
$( l^\mu_+, l^\mu_-,m^\mu)$ can then also be obtained in general (for
$\kappa\neq 0$), but they are not important here.

\end{document}